\documentclass[journal=jacsat,manuscript=article]{achemso}

\usepackage{graphicx}
\usepackage{xcolor}

\usepackage{amsfonts}
\usepackage{amsmath}
\usepackage{graphicx}
\usepackage{bm}

\newcommand{\be}{\begin{equation}} 
\newcommand{\ee}{\end{equation}}
\newcommand{\bea}{\begin{eqnarray}}
\newcommand{\eea}{\end{eqnarray}}

\title{Melting upon cooling and freezing upon heating: Fluid-solid phase diagram for
{\v S}vejk-Ha{\v s}ek model of dimerizing hard spheres}
\author{Y. V. Kalyuzhnyi}
\email{yukal@icmp.lviv.ua}
\affiliation{Institute for Condensed Matter Physics, Svientsitskoho 1, 79011 Lviv, Ukraine}
\author{P. T. Cummings}
\affiliation{Department of Chemical and Biochemical Engineering,
  Vanderbilt University, Nashville, TN 37235-1604, USA}

\begin{document}

\newpage

\begin{abstract}

A simple model of dimerizing hard spheres with highly nontrivial fluid-solid phase
behaviour is proposed. The model is studied using the recently proposed resummed 
thermodynamic perturbation theory for central force (RTPT-CF) associating potentials.
The phase diagram has the fluid branch of the fluid-solid coexistence curve 
located at a temperatures lower than those of the solid branch. This unusual behaviour 
is related to the strong dependence of the system excluded volume on the temperature, 
which for the model at hand decreases with increasing temperature. 
This effect can be also seen for a wide family of  fluid models with an effective
interaction that combines short range attraction and repulsion at a larger distance.
We expect that for sufficiently high repulsive barrier, such systems may show similar 
phase behaviour.

\begin{figure}[htbp]
\begin{center} 
\includegraphics[height=10cm,clip]{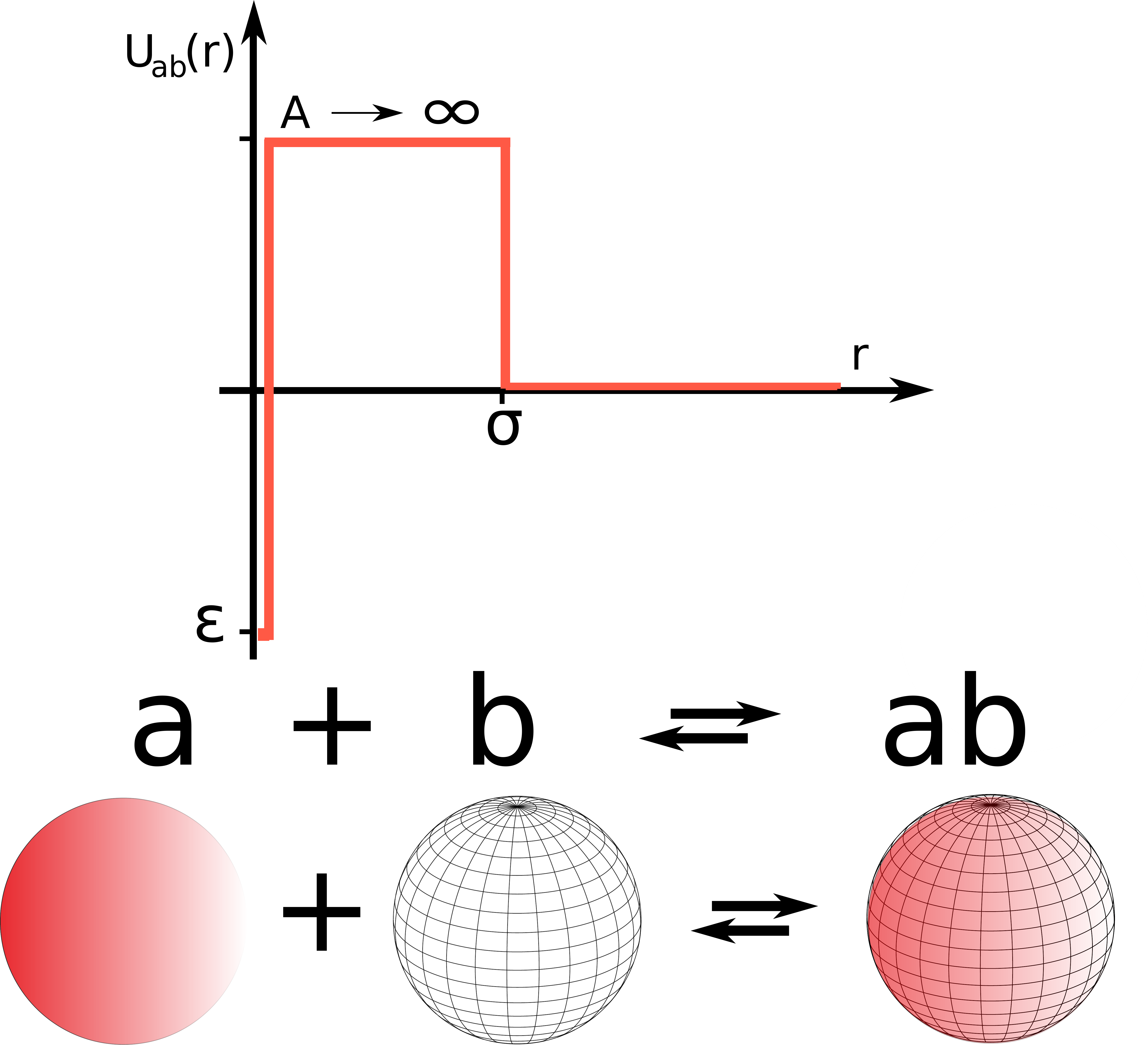}
\end{center}
\end{figure}

\end{abstract}

{\bf Keywords:} fluid-solid phase diagram, inverse melting, association, dimerization,
thermodynamic perturbation theory, phase coexistence

\maketitle

\newpage

Considerable efforts during the last decade have  focused on the investigation of 
fluid models with intermolecular pair potentials that are characterized by  
short-range attraction and long-range repulsion (SALR) (see \cite{wilding,cigala} 
and references therein). Much of the interest in studying these systems 
of this kind is due to their unusual and rich phase behaviour,
i.e. because of the competition between attraction and repulsion, the possibility exists for so-called
'modulated' phases of different type to appear. Effective interactions of the SALR type can be
observed in a number of different soft matter fluid systems, including protein solutions,
star polymers and weakly charged colloidal systems \cite{cigala,kovalchuk}. 
In addition, due to the substantial progress
obtained recently in the experimental techniques there is a possiblity to synthesize colloidal 
particles with predefined character of their interaction. A number of different versions of the
SALR models with different parameters for the shape, width and strength of the attractive and 
repulsive portions of the potential have been developed and studied recently \cite{wilding,cigala}.
These also include the shielded attractive shell (SAS) model, developed quite some time ago
\cite{cummings}. Although the SAS model was used as a simple model of chemical reaction
of dimerization, it clearly belongs to the family of SALR  models.
An important feature of these models (as also of the corresponding real systems) is that due to 
short range attractive interaction the particles form clusters. While this feature is important
 $per\;se$, for  fluids with repulsive interaction located at a distance larger than the attractive 
interaction, the formation of the clusters changes the excluded volume of the system. This effect is especially
important for the models with strong repulsion. For the low temperatures, when the degree of clusterization is
substantial, excluded volume is small. With increasing temperature, the  degree of clusterization 
becomes less and excluded volume can be substantially increased. 
This feature can have a crucial effect on the phase behaviour of the system. 

In this Letter we propose a simple SALR-type model that exhibits  so-called inverse melting,
i.e. the fluid branch of the fluid-solid phase diagram is located at lower temperatures than
the solid branch. The driving mechanism of this phase transition is related to the temperature
dependence of the excluded volume of the system.

Our model is represented by an equimolar mixture of
the hard spheres of the same size $\sigma$ with the number densities of the particles 
of the species $a$ and $b$ $\rho_a=\rho_b=\rho/2$. There is an additional square-well potential
acting between the particles of different species, so that the potentials are given by 
\be
U_{ij}(r)=U_{ij}^{(hs)}(r)+(1-\delta_{ij})U^{(ass)}(r) 
\label{Uab}
\ee
where $i$ and $j$ denote the particle species and take the values $a$ and $b$,
\be
U_{ii}^{(hs)}(r)  = 
\left\{
\begin{array}{rl}
\infty, & r<\sigma, \\
0, & r>\sigma,
\end{array}
\right.,
\;\;\;
U_{ab}^{(hs)}(r)  = 
\left\{
\begin{array}{lll}
&\infty, &\hspace{-2mm} r<L, \\
&A, \;\;\;\;\; L+\omega< &\hspace{-2mm} r<\sigma,\\
&0, &\hspace{-2mm} r>\sigma,
\end{array}
\right.
\ee
\be
U^{(ass)}(r)  = 
\left\{
\begin{array}{lll}
&0, &\hspace{-2mm} r<L, \\
&\epsilon -A,\;\;\;\;\;  L< &\hspace{-2mm} r<L+\omega,\\
&0, &\hspace{-2mm} r>L+\omega,
\end{array}
\right.,
\ee
$\epsilon$ and $\omega$ are the square-well depth
and width, respectively, and $A>0$. We will be focused here on the version of the model
with very narrow square-well potential ($\omega\ll \sigma$) and consider the limiting
case of $A\rightarrow\infty$.
The former of these conditions allows us to follow
earlier studies\cite{cummings} and approximate the square-well associating potential 
$U^{(ass)}(r)$ by the sticky potential, defined through the Boltzmann factor relationship
\be
\exp{\left[-\beta U^{(hs)}_{ab}(r)\right]}\left\{
\exp{\left[-\beta U^{(ass)}(r)\right]}-1\right\}
={\sigma^3\over 12L^2\tau}\delta(r-L),
\label{stick}
\ee
where $\tau$ is Baxter's stickiness parameter \cite{Baxter1968}. 
The relation between stickiness parameter
$\tau$ and temperature $\beta$ can be established by equating the second virial 
coefficient of the limiting sticky case (Eq. \ref{stick}) with that of the original model defined in Eqs. 1-3.
In particular, we will consider here the version of the model with the square-well placed in a center 
of the sphere, i.e. $L\rightarrow 0$. Thus upon association the 
model can form only dimers with one hard sphere completely buried inside the other 
hard sphere. We will refer to this model as {\v S}vejk-Ha{\v s}ek 
({\v S}H) model \cite{svejk} of dimerizing hard spheres.

The theoretical description of the model at hand can be carried out using the recently proposed
extension of the resummed thermodynamic perturbation theory for central force associating
potentials (RTPT-CF) \cite{rescic2016}. The important feature of the model is
that at any temperature the system can be treated as a one-component hard-sphere fluid
with effective density 
\be
\rho_{eff}={1\over 2}(\rho+\rho_0), 
\label{eff}
\ee
where $\rho_0$ is the total density
of nonbonded particles.
This feature enables us to calculate the fluid-solid phase coexistence diagram of the 
{\v S}H model using the phase diagram of the conventional hard-sphere model.
We shall denote the coexisting densities of such phase diagram as
$\rho_{F}^{(0)}$ and $\rho_{S}^{(0)}$, 
where the lower indices $F$ and $S$ denote the fluid and solid phases, respectively.

According to RTPT-CF approach \cite{rescic2016} the density of nonbonded 
particles $\rho_{0}$ satisfies the following equation:
\be
{1\over 2}E\rho_{0}^2+\rho_{0}-\rho=0,
\label{eq}
\ee
where 
\be
E={\pi\sigma^3\over 3\tau}y_{hs}(0;\rho_{eff})
\label{E_c}
\ee
and $y_{hs}(0;\rho_{eff})$ is the cavity distribution function
of the hard-sphere fluid at the overlapping distance $r=0$ and
with the particle number density $\rho_{eff}$.
This cavity distribution function can be calculated using relation due to Hoover and Poirier 
\cite{hoover1962}. For the model at hand we have:
\be
\ln{\left[y_{hs}(0;\rho_{eff})\right]}=
\beta\mu^{(ex)}_{hs}(\rho_{eff}).
\label{ln_y}
\ee
Taking into account the expression for  $\rho_{eff}$ (\ref{eff}) equation 
(\ref{eq}) can be used to calculate the effective density of the system 
 as a function of the temperature $\tau$ and density $\rho$.
For $\rho_{eff}$ equal to either $\rho_F^{(0)}$ or $\rho_S^{(0)}$ equation (\ref{eq})
couples the temperature $\tau$ and density $\rho$ along either fluid or solid branches
of the fluid-solid phase diagram of our {\v S}H dimerizing model. In this case we have
\be
\rho={2\left[1+\rho_K^{(0)}E(\tau,\rho_K^{(0)})-
\sqrt{1+\rho_K^{(0)}E(\tau,\rho_K^{(0)})}\right]\over E(\tau,\rho_K^{(0)})},
\label{branch}
\ee
where
\be
E(\tau,\rho_K^{(0)})={\pi\sigma^3\over 3\tau}
\exp{\left[\beta\mu_{hs}^{(ex)}(\rho^{(0)}_K)\right]}
\label{muK}
\ee
and the subscript $K$ takes the value of either $F$ or $S$.
The excess chemical potential $\mu^{(ex)}_{hs}(\rho^{(0)}_{F})$, which is needed to calculate 
liquid branch of the phase diagram, can be obtained using the Carnahan-Starling expression 
\cite{Carnahan1969}. 
At equilibrium, the chemical potential in both phases is equal. Hence, 
the excess chemical potential in the solid phase 
$\mu^{(ex)}_{hs}(\rho^{(0)}_{S})$ can be calculated using its value in the fluid phase, 
i.e.
\be
\beta\mu^{(ex)}_{hs}(\rho^{(0)}_{S})=
\beta\mu^{(ex)}_{hs}(\rho^{(0)}_{F})+\ln{\rho_{F}^{(0)}}-
\ln{\rho_{S}^{(0)}}.
\label{muS}
\ee

Using the relation between $\rho$ and $\tau$ (\ref{branch}), along with 
with computer simulation values of the hard-sphere fluid-solid coexisting densities
(i.e., $\rho_F^{(0)}\sigma^3=0.9427$ and $\rho_S^{(0)}\sigma^3=1.0411$ \cite{Hoover1968}), 
the fluid-solid phase diagram for the {\v S}H model of dimerizing hard spheres was computed. 
The corresponding phase diagram in $\tau$ vs $\rho$ coordinate frame is presented
in figure 1. In contrast to conventional fluid-solid phase diagram one can see that the
fluid branch of the diagram is located at lower temperatures than the solid branch. Thus at constant
density, with increasing temperature  the system undergoes a fluid to solid phase transition
(and vice versa). At sufficiently low temperature all the particles will be dimerized and
the effective density of such system will be $\rho_{eff}=\rho/2$. If in this case
$\rho < 2\rho_F^{(0)}$ then the system will be in a fluid state. With increasing temperature, the number of dimers decreases and the effective density increases. As soon as 
$\rho_{eff}$ becomes equal $\rho_F^{(0)}$, the system begins to freeze and upon reaching
the value of $\rho_S^{(0)}$ the system will be in a solid state. According to Figure 1 this
may happen if the density of the system is in the range $\rho_S^{(0)}<\rho<2\rho_F^{(0)}$.
A system with density $\rho_F^{(0)}<\rho<\rho_S^{(0)}$ and sufficiently high temperature
will be in a state of fluid-solid equilibrium with the densities of the coexisting phases
$\rho_F^{(0)}$ and $\rho_S^{(0)}$. Similarly, for  density in the range
$2\rho_F^{(0)}<\rho<2\rho_S^{(0)}$ and sufficiently low temperature the system will 
split into the coexisting fluid and solid phases with the densities $2\rho_F^{(0)}$ 
and $2\rho_S^{(0)}$, respectively.

In summary, we have studied a simple model for dimerization with highly nontrivial fluid-solid phase
behaviour. The phase diagram, which was build using recently proposed RTPT-CF, has the 
liquid branch of the coexisting curve located at a temperatures lower than those of the 
solid branch. This unusual behaviour is related to strong dependence of the system
excluded volume on the temperature, which for the model at hand decreases with increasing temperature. 
This effect can be observed also for most of the SALR type of the models and we expect that
 systems with sufficiently high repulsive barrier may exhibit similar phase behaviour.

%{\bf Acknowledgment}

\begin{figure*} 
\begin{center}
\includegraphics[width = 1.0\textwidth]{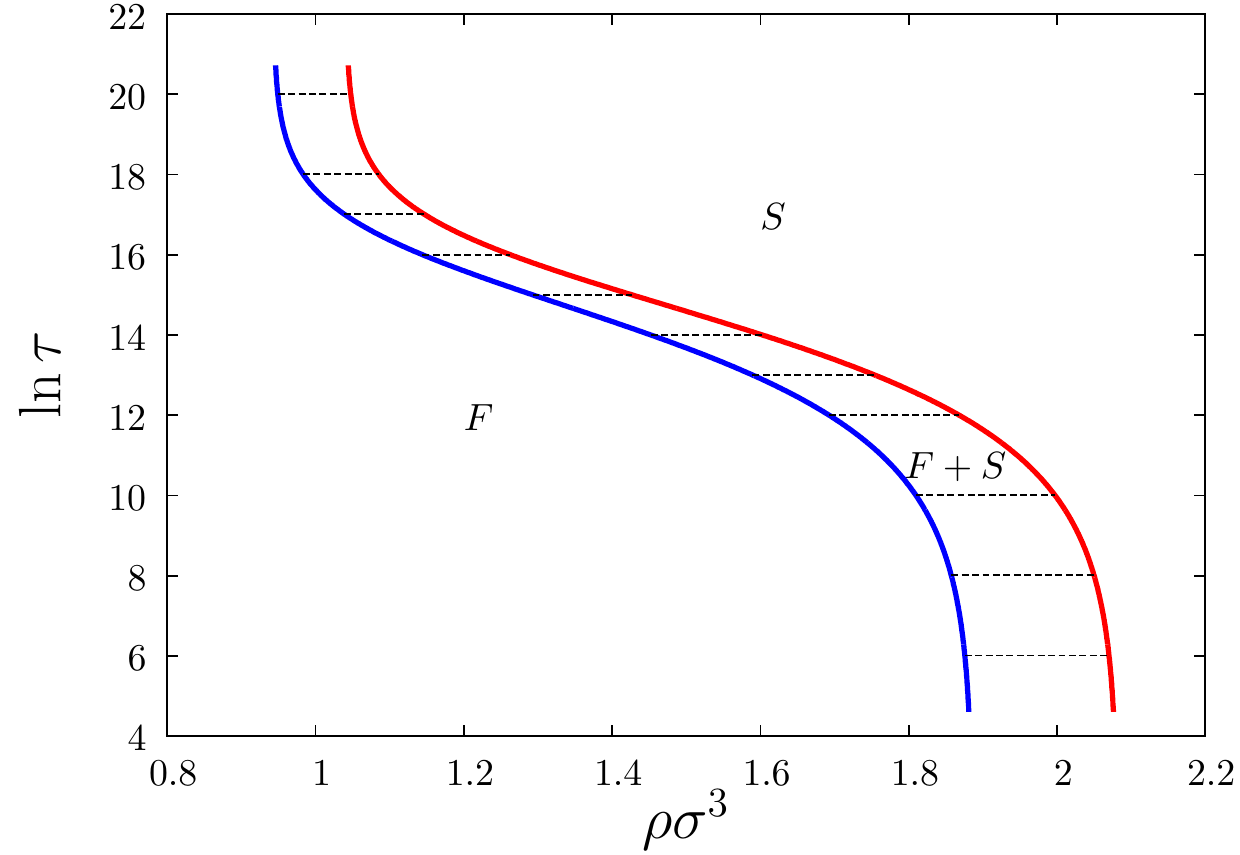}
\caption{Fluid-solid coexistence diagram for the {\v S}H model of dimerizing hard spheres in 
$\ln{\tau}$ vs $\rho\sigma^3$ coordinate frame. The fluid-solid phase boundaries are denoted
by the solid lines with blue and red lines standing for the fluid and solid branches, respectively. 
Here black dashed lines denote tie lines, $F$ denote the fluid phase, 
$S$ denote the solid phase and $F+S$ denote the region of fluid-solid
coexistence.}
\label{fig1}
\end{center}
\end{figure*}

\end{document}